\documentclass[preprint,aps,prb,longbibliography]{revtex4-1}
\usepackage[dvipdfmx]{graphicx}
\usepackage{dcolumn}
\usepackage{bm}

\usepackage{color}

\usepackage{amsmath}
\usepackage{amsfonts}
\usepackage{amssymb}
\providecommand{\U}[1]{\protect\rule{.1in}{.1in}}

\begin{document}

\title{Local polarization and valence distribution in LaNiO$_3$/LaMnO$_3$ heterostructures}

\author{Masato Anada$^{1,2}$, Satoshi Sakaguchi$^1$, Kazuki Nagai$^{1,2}$, Miho Kitamura$^3$, Koji~Horiba$^3$, Hiroshi Kumigashira$^{3,4,5}$, and Yusuke Wakabayashi$^{2,5}$\footnote{wakabayashi@tohoku.ac.jp}}

\affiliation{
$^1$Division of Materials Physics, Graduate School of Engineering Science, Osaka University, Toyonaka 560-8531, Japan\\
$^2$ Department of Physics, Graduate School of Science, Tohoku University, Sendai 980-8578, Japan\\
$^3$Institute of Materials Structure Science, High Energy Accelerator Research Organization, Tsukuba 305-0801, Japan\\
$^4$Institute of Multidisciplinary Research for Advanced Materials, Tohoku University, Sendai 980-8577, Japan\\
$^5$ Materials Research Center for Element Strategy, Tokyo Institute of Technology, Yokohama 226-8501, Japan.}

\date{\today}

\begin{abstract}
The inside of the electrical double layer at perovskite oxide heterointerfaces is examined. Here, we report the local polarization and valence distribution in LaNiO$_3$/LaMnO$_3$ and LaMnO$_3$/LaNiO$_3$ bilayers on a SrTiO$_3$ (001) substrate. Simultaneous measurements of two aspects of the structure are realized by using Bayesian inference based on resonant- and nonresonant-surface X-ray diffraction data. The results show that the average Mn valences are Mn$^{3.12+}$ and Mn$^{3.19+}$ for the two samples. The intensity of their local electric field is $\sim$1~GV/m and the direction of the field points from LaMnO$_3$ to LaNiO$_3$.

\end{abstract}

\maketitle

\section{Introduction}

The epitaxial interfaces of strongly correlated oxides are where electrons having different states meet each other\cite{Hwang12NatureMat,Chakhalian14RevModPhys,Bhattacharya14AnnualRevMatRes,Middey16AnnualRevMatRes}. The behavior of electrons at these interfaces is different from that in the bulk material because electrons will interact with those on the other side of the interface. 
Various interfacial properties can be interpreted by electron transfer across the interface\cite{Hoffman13PRB,Kitamura16APL,Wei17APL} and self doping\cite{Wang15Science}, which originate from the local electric field or built-in potential. The local electric field is induced to achieve a proper band alignment\cite{Giampietri17AdvMatInt,Zhong17PRX}. Because of the high density of the electric charge in oxide materials, the spatial distribution of the potential gradient in oxide heterostructures is often limited to a few nanometers, which results in strong electric field. According to theoretical calculations\cite{Dorin19NewJPhys,Chen17PRL} and cross-sectional measurements\cite{Huang12PRL}, some transition metal oxide interfaces have a potential gradient of $\sim$1~GV/m. The spatial distribution and magnitude of the electric field differ by several orders of magnitude from those in classical semiconductors, in which a spatial distribution of 10~nm to 100~nm and local electric field of $\sim$ MV/m are observed.

 In the case of ABO$_3$ perovskite oxides, AO termination and BO$_2$ termination causes a stark difference in transport properties around the surface\cite{Ohtomo04Nature,Nakagawa06NatMat,Kumah14PRAppl}. The termination also affects the work function\cite{Jacobs16AdvFunctMat,Zhong16PRB}, which is directly related to the band alignment. 
 Even if we have a technique to arrange the atoms as designed, an interface or surface reconstruction of atomic\cite{Herger07PRB,Kan14JAP,Yuan18NatureCommun,Fowlie19NanoLett}, valence\cite{Grisolia16NaturePhys,Chandrasena18PRB,Zabolotnyy18PRB}, and orbital arrangements\cite{Wakabayashi07NatMat,Benckiser11NatureMat,Park13PRL} occur spontaneously, and the local field is altered. Therefore, experimental observation of the interfacial valence- and atomic-arrangements is needed to understand and control the oxide interface properties.
So far, the spatial distribution of the valence around the interface has been examined by cross-sectional electron energy-loss spectroscopy measurements\cite{Ohtomo02Nature,Gibert15NanoLett} or the capping layer thickness dependence of the X-ray absorption spectroscopy (XAS) spectra\cite{Kitamura15APL,Kitamura16APL,Kitamura19PRB}. These methods provide the spatial distribution of elements and their valences, while being insensitive to the electric polarization.

 In this study, we have investigated the interfacial structure of a set of typical heterointerfaces, a LaNiO$_3$ (LNO) on LaMnO$_3$ (LMO) bilayer and an LMO on LNO bilayer that are grown on top of SrTiO$_3$ (STO) (001) surfaces (hereafter, we shall refer to these as NMT and MNT, respectively). The LNO/LMO interface is known as a typical paramagnetic/ferromagnetic interface, and the interfacial magnetic interaction, which includes the exchange bias effect, has been studied extensively\cite{Gibert12NatMat,Piamonteze15PRB,Kitamura16APL,Wei17APL,Kitamura19PRB}. 
 We used the crystal truncation rod (CTR) scattering method, which is one of the surface X-ray diffraction methods, in this study. The advantage of using this technique is its high resolution that allows us to examine the local polarization as well as to study interplane distances\cite{Willmott07PRL,Yamamoto11PRL,Kumah14AdvMat,Anada18PRB}. In addition, we also measured the Mn K-edge energy spectra of CTR scattering\cite{Chu99PRL,Perret13JApplCryst,Joly18JChemTehoryComp,Grenier18PRB} to determine the spatial distribution of Mn$^{4+}$ ions. 
Our resonant surface X-ray diffraction measurements reveal the interfacial valence distribution as well as the atomic arrangement at the sub-\AA\/ spatial resolution. 

\section{Experiment and analysis method}

An LMO and LNO (LNO and LMO) heterojunction, with a thickness of $5+5$ unit cells, was deposited on an atomically flat (001) surface of a Nb-doped STO substrate by pulsed laser deposition (PLD). During deposition, the substrate temperature was controlled at 500~$^\circ$C and 600~$^\circ$C for LNO and LMO growth under an oxygen pressure of $1\times10^{-3}$ Torr. The film thickness was precisely controlled on the atomic scale by monitoring the intensity oscillation of the reflection high-energy electron diffraction. After the deposition, the samples were post-annealed in 760~Torr oxygen at 400~$^\circ$C for 45 minutes to fill the oxygen vacancy in LNO. In LMO, however, excess oxygen ions were expected to be introduced by this post annealing process. 

The CTR scattering measurements were performed with the four-circle diffractometer installed at BL-3A, Photon Factory, KEK, Japan. The scattered X-ray intensity was measured by a two-dimensional pixel array detector XPAD-S70, imXpad, France. Samples were placed in a 10$^{-5}$ Torr vacuum chamber to protect them from radiation damage, and all the measurements were performed at room temperature. Using 12~keV X-rays, which is a non-resonant condition, the $hk\zeta$ rods along ($h$,$k$) = (0,0), (0,1), (0,2), (1,1), and (1,2) were measured. The range of measured $\zeta$ was 0.5 to 4.2 except for the substrate Bragg points. Additionally, $00\zeta$ profiles were measured using several X-ray energies ($E$) close to the Mn and Ni $K$-absorption edges to discriminate B-site elements. To study the spatial distribution of Mn$^{4+}$, the energy spectra around the Mn $K$-edge were obtained for 6 (NMT) and 12 (MNT) scattering vectors along the $00\zeta$ line. 

After the measurements, illumination area correction and Lorentz factor correction were performed by following the procedure described in Ref.~\onlinecite{Schleputz05ActaCrystA}.
The intensity profiles along the $hk\zeta$ rods measured in both non-resonant and resonant conditions were simultaneously analyzed by exchange Monte Carlo (MC) sampling software based on Bayesian inference\cite{Anada17JApplCryst,Nagai19JApplCryst}. Both LNO and LMO were treated to have a pseudocubic unit cell. In our analysis, the in-plane atomic positions were fixed to the simple cubic SrTiO$_3$ structure. The other parameters, i.e., atomic displacement in the surface normal direction with respect to the substrate lattice ($\delta z$), atomic occupancy (occ), and isotropic atomic displacement parameters ($B_{\mathrm{iso}}$), were refined. The total atomic occupancy for each site was fixed to unity except for the surface region. The total number of structure parameters was 90. The parameter uncertainty was estimated using MC sampling by taking into account the effective number of observations\cite{Nagai19JApplCryst}.

The spatial distribution of Mn$^{4+}$ was derived from the energy spectra measured at several scattering vectors. 
The spectra depend on the non-resonant scattering amplitude as well as the spatial distribution of Mn$^{4+}$. Although the non-resonant amplitude was calculated using the result of the interfacial structure analysis described above, it inevitably contains some error. The error was corrected by adding a small complex value to the non-resonant scattering amplitude. 
The anomalous scattering factors $f'$ and $f''$, as functions of $E$ for Mn$^{3+}$, were calculated by the finite difference method near edge structure (FDMNES) code\cite{Bun09JPhysCondenMatt} using the bulk LaMnO$_3$ structure\cite{Rodriguez-Carvajal98PRB}. The anomalous scattering factors for Mn$^{4+}$ were obtained by applying a chemical shift of 4~eV \cite{Nakao09JPSJ} from those for Mn$^{3+}$.
The analysis was also performed with the Bayesian inference by assuming that the magnitude of the noise was 10~\% of the intensity.
 The error bars of the Mn$^{4+}$ concentration were estimated by MC sampling, while the effective number of observations of the energy spectra was assumed to be the same as the number of data points.

\section{Results}

\begin{figure*}
	[tb]
	\begin{center}
		\includegraphics[width=16cm]
		{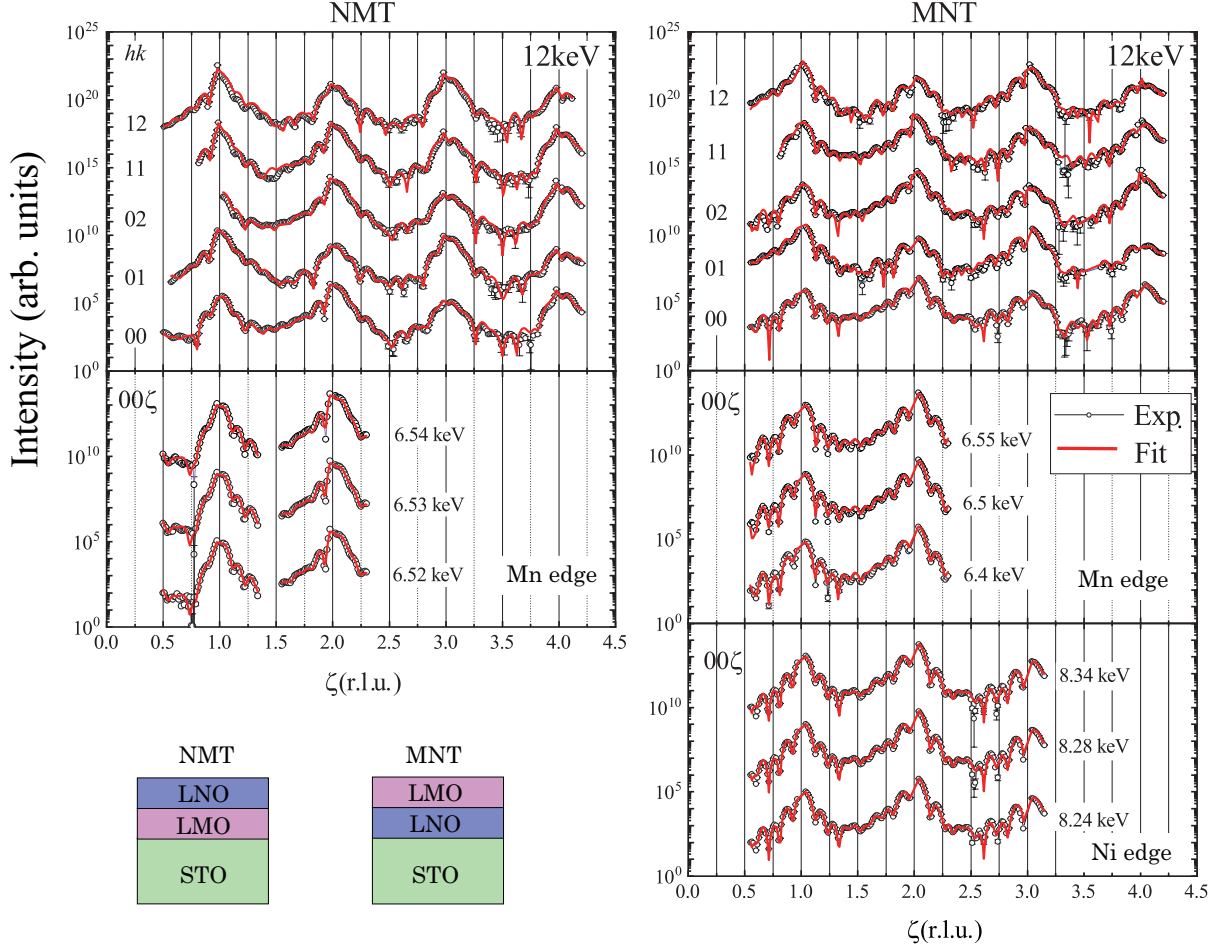}
	\end{center}
	\caption{CTR scattering profiles along the $hk\zeta$ rods for NMT (left) and MNT (right). The middle and bottom panels show the $00\zeta$ rods in a resonant condition close to the Mn and Ni $K$-absorption edges. Open symbols show the experimental results, and the solid curves show the calculated intensity derived from the refined structure. The error bars show the statistical error.}
\label{fig:1}
\end{figure*}

\begin{figure*}
	[tb]
	\begin{center}
		\includegraphics[width=15.5cm]
		{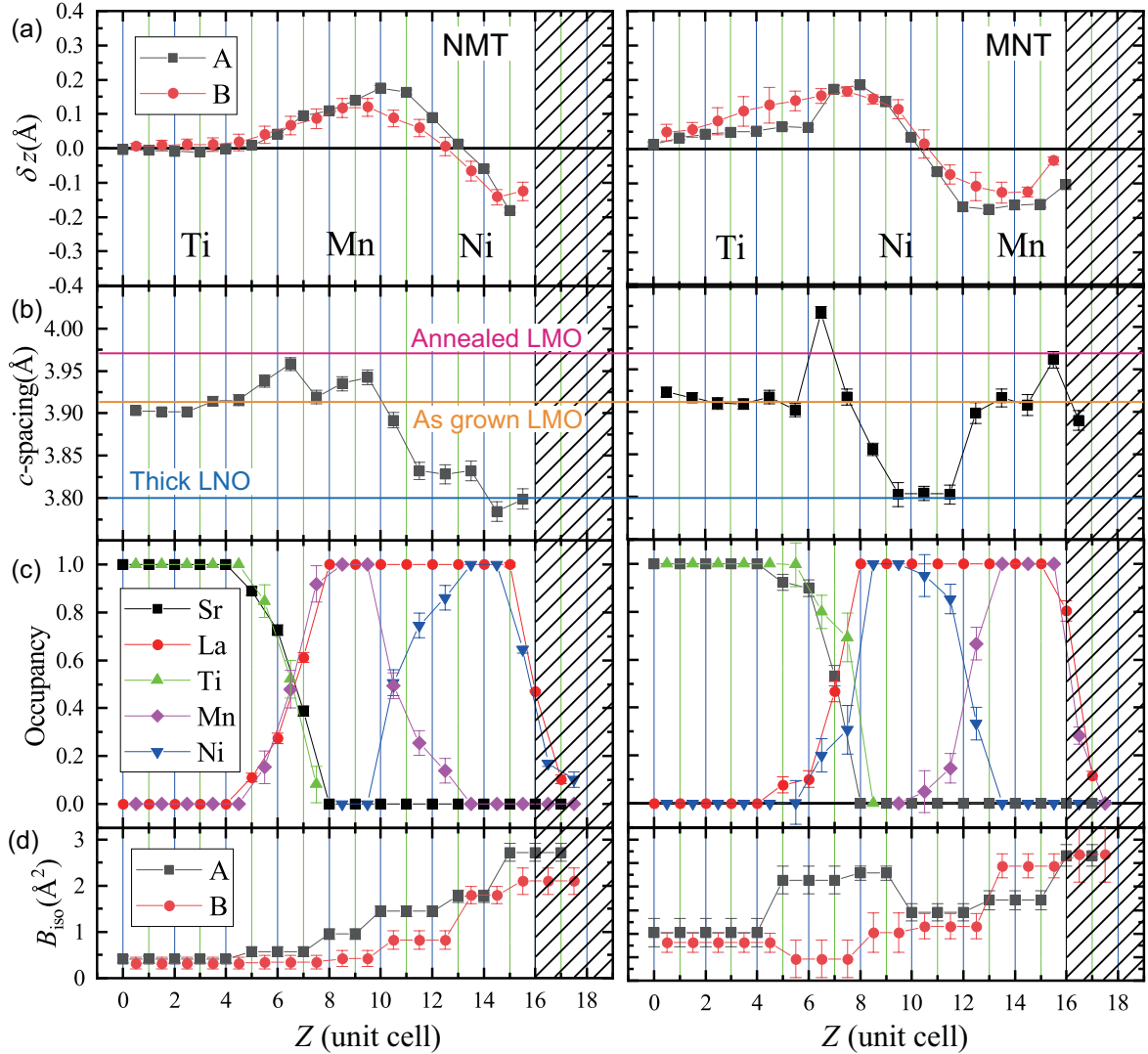}
	\end{center}
	\caption{(a) A-site and B-site atomic displacements in the outward surface normal direction from the SrTiO$_3$ substrate lattice, (b) $c$-lattice spacing defined by the A-site distance, (c) atomic occupancy of A-site and B-site atoms, and (d) atomic displacement parameters as a function of depth. The horizontal lines in panel (b) show the interplane distances of the thick films of (blue) LNO\cite{May10PRB,Sanchez00ApplphysA}, (orange) as-grown ferromagnetic metal LMO, and (magenta) annealed canted magnetic insulator LMO\cite{Choi09JPhysD}.}
\label{fig:2}
\end{figure*}

Non-resonant and resonant CTR profiles for MNT and NMT films are presented in Fig.~\ref{fig:1}. Open symbols show the experimental results and the solid curves show those obtained from calculation. The $R$ ($\equiv \sum^{}_{\zeta ,E}[|(|F_{\rm expt}|-|F_{\rm calc}|)|]/[\sum^{}_{\zeta ,E}|F_{\rm expt}|]$) values, where $F_{\rm expt}$ and $F_{\rm calc}$ denote the square root of the measured and calculated intensities, respectively, were 0.11 for MNT and 0.10 for NMT. The resulting structural parameters, $\delta z$, occ, and $B_{\mathrm{iso}}$ for each site, are shown in Fig.~\ref{fig:2} as functions of depth, $Z$. The surface was $Z\simeq 16$ and the STO substrate corresponded to the $Z<4$ region. 

The $c$-lattice spacing is defined by the slope of $\delta z$. Deep inside the substrate, the slope is zero by definition. The $c$-lattice spacings for NMT (MNT) were 3.827$\pm0.008$~\AA\/ (3.803$\pm$0.009~\AA\/) in the LNO region, and 3.938$\pm$0.007~\AA\/ (3.91$\pm$0.01~\AA\/) in the LMO region. Figure~\ref{fig:2} (b) shows the depth profile of the lattice spacing defined by the interatomic distance of the A-site cations.

Using the atomic occupancy around the interfaces, one can evaluate the degree of atomic interdiffusion. Reference \onlinecite{Gibert15NanoLett} reports more Mn-Ni interdiffusion in LNO on LMO interfaces than in LMO on LNO interfaces for sputter-made samples, and Ref. \onlinecite{Kitamura16APL} reports sharp interfaces on both sides for PLD-made samples. Fig.~\ref{fig:2} (c) shows only a small difference in atomic interdiffusion between the two PLD-made samples. In addition, we observe LaTiO$_3$ formation at the STO/LNO interface in the MNT sample. The LaTiO$_3$ layer had a large $c$-lattice spacing, as shown in panel (b), which reflected its large cell volume\cite{Okamoto06PRL}. At the same position, the value of $B_{\mathrm{iso}}$ for the La ions was found to be large compared with that for the B-site ions. This should be the signature of the different octahedral rotation mode around the LaTiO$_3$ layer. Unlike cubic STO or rhombohedral LNO, bulk LaTiO$_3$ exhibits a GdFeO$_3$-type structure with La displacements\cite{Cwik03PRB}. 

\begin{figure*}
	[tb]
	\begin{center}
		\includegraphics[width=16cm]{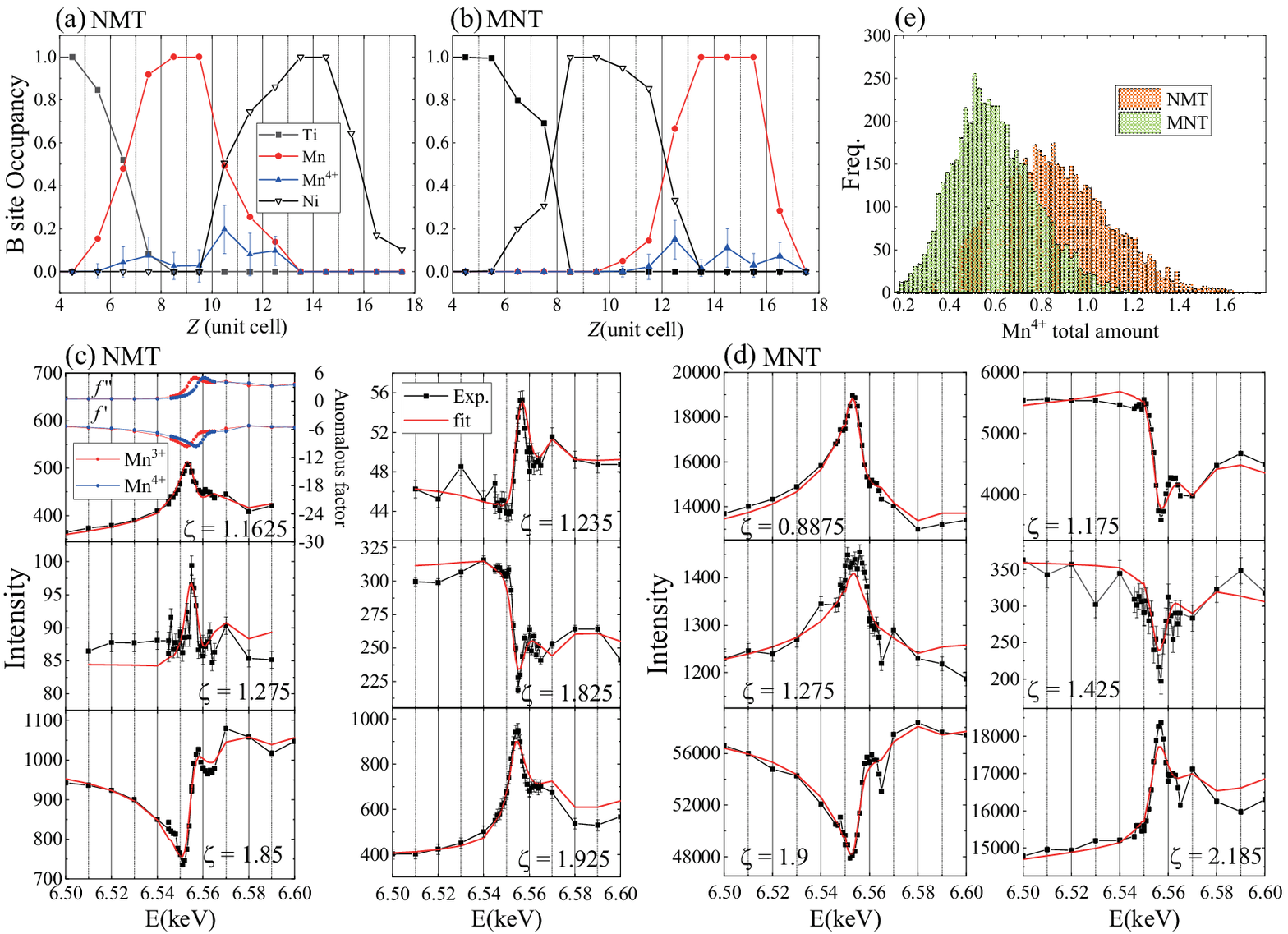}
		\caption{(a) Mn$^{4+}$ occupancy as a function of depth for NMT, together with the occupancies of three kinds of B-site elements. (b) Mn$^{4+}$ occupancy as a function of depth for MNT, together with the occupancies of three kinds of B-site elements. (c) Energy spectra of the scattered X-ray intensity around the Mn $K$-absorption edge measured at several points on the $00\zeta$ rod for NMT. Black plots show the experimental results and red solid lines show the calculated intensities. Anomalous scattering factors $f'$ and $f''$ for Mn$^{3+}$ and Mn$^{4+}$ used for the analysis are also shown in the panel for $\zeta=1.1625$. (d) Energy spectra of the scattered X-ray intensity around the Mn $K$-absorption edge measured at several points on the $00\zeta$ rod for MNT. The colors are the same as in (c). (e) Probability density of the total amount of Mn$^{4+}$ derived from the energy spectra.}
\label{fig:3}
	\end{center}
\end{figure*}

\begin{figure}
	[tb]
	\begin{center}
		\includegraphics[width=8cm]{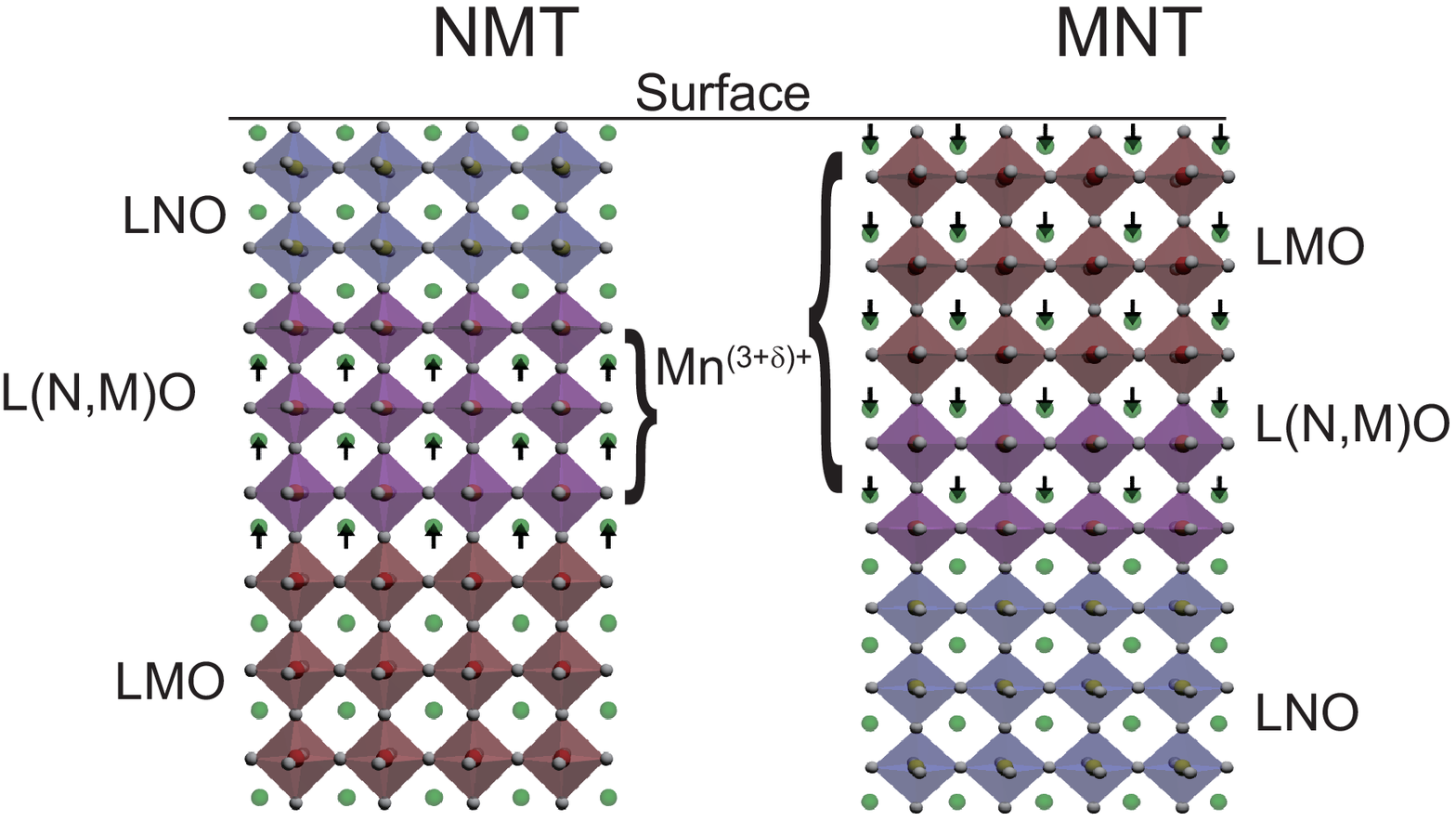}
		\caption{Schematic view of the analyzed interfacial structures of the two samples. Blue and red octahedra represent NiO$_6$ and MnO$_6$, and purple octahedra show (Ni,Mn)O$_6$ octahedra. The Mn$^{4+}$ distribution is also marked. The green spheres show the La ions and the arrows on them show the direction of the atomic displacement caused by the local polarization.}
\label{fig:summary}
	\end{center}
\end{figure}

Energy spectra around the Mn absorption edge measured at several points on the $00\zeta$ rod are shown in Fig.~\ref{fig:3} (c) and (d). The MC calculation derived the spatial distribution of Mn$^{4+}$, as shown in panels (a) and (b). As can be seen, the Mn$^{4+}$ was concentrated at the interface in NMT, while the MNT sample had a rather homogeneous Mn$^{4+}$ distribution in the entire LMO region. At the surface and interface of LMO, the hole concentration was increased even if the Mn$^{4+}$ occupancy was homogeneous because the occupancy of the Mn was reduced. To obtain the total amount of electron transfer from Ni to Mn, the occurrence distribution of the total amount of Mn$^{4+}$ in the MC calculation is plotted in panel (e).
 The amount of Mn$^{4+}$ in NMT (MNT) was $0.8\pm0.3$ $(0.5\pm0.2)$. 
Total number of Mn layers for NMT and MNT, derived by the same method, were 4.4~$\pm$~0.2 and 4.2~$\pm$~0.1, and thus the average hole concentrations of the Mn site in the two samples were 19$\pm$8~\% and 12$\pm$5~\%, respectively. The observed interfacial structure and the Mn$^{4+}$ distribution for the two samples are schematically summarized in Fig.~\ref{fig:summary}.

\section{Discussion}

\subsection{Valence distribution}

The total amount of Mn$^{4+}$, which is a measure of the electron transfer across the LMO/LNO interface in NMT (MNT), was obtained as $0.8\pm0.3$ $(0.5\pm0.2)$, as shown in Fig.~\ref{fig:3} (e). These values are consistent with the previously reported values based on XAS measurements\cite{Kitamura19PRB}.

In the LNO region where the Ni occupancy is more than 80~\%, the $c$-spacings for both samples were very close to that of thick LNO films on STO\cite{May10PRB,Sanchez00ApplphysA}, 3.80~\AA\/, except for the surface or interface layers. This means that the LNO growth is well controlled and stable.

Next, we focused our attention on the LMO region where the Mn occupancy is more than 80~\%.
The interplane distance in LMO is known to relate to the hole concentration $x$ and, as its natural consequence, to the magnetic and transport properties.
 Most of LMO films grown on STO show ferromagnetism, which is different from the canted magnetism of the bulk LMO. Reference \onlinecite{Choi09JPhysD} reported the magnetism, electric property, and $c$-lattice spacing of a 30~nm-thick PLD-made LMO film on a STO (001) substrate. The as-grown film showed ferromagnetism, and oxygen-reducing atmosphere annealing made the property of the LMO film bulk-like. The reported $c$-lattice spacing of the as-grown and annealed samples were 3.912~\AA\/ and 3.973~\AA\/, respectively, and these are shown in Fig.~\ref{fig:2} (b) with horizontal lines. 
The lattice spacings of LMO in our NMT and MNT samples were 3.94~\AA\/ and 3.91~\AA\/, respectively. The quantitative similarity of the LMO lattice spacing of the MNT sample and that of as-grown LMO is apparent. The spacing in NMT sample had the intermediate value of the as-grown LMO and annealed LMO. The difference between NMT and MNT was caused by the successive film growth on top of LMO, which may work as an oxygen reduction annealing. 
The valence of the Mn ions is derived from the $c$-lattice spacing. For this purpose, the (La,Ca)MnO$_3$ system\cite{Hibble99JPhysCondMat,Ganguly00JPhysCondMat} was selected as the reference because the ionic radii of La and Ca are similar. The cell volume of the La$_{1-x}$Ca$_x$MnO$_3$ was 242$-$41$x$~\AA\/$^3$. Using the Poisson ratio of LMO of 0.3\cite{osti_1193810}, we obtained the hole concentration dependence of the $c$-lattice spacing $dc/dx$=$-$0.34~\AA\/, i.e., one expects a $c$-lattice spacing contraction of 0.34~\AA\/ when oxidizing LaMn$^{3+}$O$_3$ to LaMn$^{4+}$O$_3$.
Here, we selected the $c$-lattice spacing of the annealed LMO as the reference value for Mn$^{3+}$. In this case, the $c$-lattice spacings for the as-grown LMO\cite{Choi09JPhysD}, and the LMO region in NMT and MNT corresponded to Mn$^{3.18+}$, Mn$^{3.12+}$, and Mn$^{3.19+}$, respectively. 
The estimated valences from the lattice spacing showed clear similarity with those estimated from the energy spectra shown in Fig.~\ref{fig:3}.

We then examined the spatial distribution of the valence at the interface region.
At the LNO/LMO interface, Ni$^{2+}$ has been reported by XAS measurements\cite{Kitamura16APL} for a thickness of only one u.c. 
In the present study, the lattice spacing of LNO was approximately 3.80~\AA\/ for both samples, which is the same as the value for a thick LNO film on STO\cite{May10PRB,Sanchez00ApplphysA} within the experimental uncertainty even at the LNO/LMO interface.
Similar structural and spectral features have been reported at the LNO/STO interface\cite{Tung17PhysRevMat}. In their report, the introduction of Ni$^{2+}$ with little volumetric change was attributed to the oxygen vacancy stabilization caused by 
the tensile strain on LNO caused by the STO substrate. 
Note that the cell volume of LMO is nearly the same as that of STO.

Next, the possible distribution of Mn$^{2+}$ at the LMO/STO interface, which was reported by XAS\cite{Chen17PRL}, was examined. 
The Shannon ionic radius of Mn$^{2+}$ is larger than that of Mn$^{3+}$ by 0.18~\AA. This difference in ionic radius between 2+ and 3+ ions is larger than the difference between 3+ and 4+ ions (0.12~\AA\/), and therefore Mn$^{2+}$ affects the lattice spacing as much as Mn$^{4+}$. Thus, we assumed as constant $dc/dx$, $-$0.34~\AA\/, for the $-1<x<1$ range. 
 Figure~\ref{fig:2} (b) shows that there was an elongation of 0.04~\AA\/ for the $c$ lattice spacing at the $Z=6.5$ position (LMO/STO interface). The amount of interfacial expansion corresponded to the addition of 0.12 electrons/Mn ion. In the LMO region for NMT, the average valence of Mn was estimated to be Mn$^{3.12+}$. Therefore, the in-plane average of the Mn valence at the $Z=6.5$ position  was close to Mn$^{3+}$. This result means that there should be as much Mn$^{4+}$ as Mn$^{2+}$ at the interface.

The spatial distribution of the valence of Mn around the LNO/LMO interface derived from the energy spectra is shown in Fig.~\ref{fig:3}(a) and (b). The Mn$^{4+}$ ions distributed in a wide range in the MNT sample, and were concentrated at the LMO/LNO interface in the NMT sample for a 2- to 3-u.c.-thick region. The region where Mn$^{4+}$ was concentrated in the NMT sample had a Ni occupancy of more than 70~\% and a small $c$-lattice spacing of 3.83~\AA\/, which is very close to the lattice spacing of LNO. The Mn ions mixed into the LNO side had little volume, and the pressure made Mn$^{4+}$ stable. 

\subsection{Local polarization}
The local electric polarization was estimated from the depth profile of $\delta z$ presented in Fig.~\ref{fig:2}(a). The local electric polarization is a good measure of the local electric field because LMO is a paraelectric material.
 The electric polarization in perovskite oxides is mainly caused by the Slater mode and the Last mode atomic displacements, and these both involve a relative displacement of A-site and B-site ions. In the case of compounds whose tolerance factor is smaller than 1, B-site ions have little room around them whereas A-site ions have more room, and the Last mode is the main origin of the polarization. Both LNO and LMO are assigned to this class of compounds. 
Theoretical calculation\cite{Smirnova99PhysicaB} on the $Pnma$ phase of LaMnO$_3$ actually showed that the three lowest energy modes of $B_{u}$ symmetry, which correspond to the electric polarization parallel to the $a$, $b$ and $c$ axes, were mainly composed of La displacements. Therefore, one can approximate the lattice system as fixed BO$_6$ octahedra and loosely bound A-site ions. 
The typical energy of the lowest energy modes\cite{Smirnova99PhysicaB} is 90~cm$^{-1}$, which means the force constant for La displacement is 70~N/m. Using this force constant, one can estimate the intensity of the local electric field at the La position.

The LNO regions in both samples showed little relative displacement of A- and B-site ions, which meant that there was no polarization. No polarization was as expected because LNO is metallic.
A large polarization was observed at the interface between LMO and LNO in the NMT sample and at the LMO region in the MNT sample. The polarization in the two samples showed opposite directions. The local electric field in the LMO region pointed towards the LNO layer.
The observed relative displacements of A-site and B-site ions in NMT was 0.08$\pm 0.01$~\AA\/, and that in MNT was 0.05$\pm 0.02$~\AA\/ for 2 to 3 layers around the interface. Using the valence of La as $3+$, we obtained the intensity of the local electric field as 1 (0.7)~GV/m pointing from LMO to LNO in the NMT (MNT) sample.

Qualitatively, the direction of the electric field is consistent with the band bending to align the oxygen $2p$ level\cite{Zhong17PRX}. 
 Quantitatively, this estimation of the electric field intensity involves an uncertainty of $\sim$50~\%, because the lowest energy $B_u$ modes range from 76 to 90~cm$^{-1}$, and the Born effective charge of the rare-earth ion $R^{3+}$ for $R$MnO$_3$\cite{Vermette12PRB,Jandl13JPhysCondMat,Bukhari16ActaPhysica} ranges from 3.3+ to 3.8+ instead of 3+. 
 The obtained magnitude of the electric field is similar to the values reported for various perovskite heterostructures based on tunneling transport measurement\cite{Singh-Bhalla11NaturePhys}, cross-sectional scanning tunneling microscopy\cite{Huang12PRL}, and first-principles calculations\cite{Chen17PRL,Dorin19NewJPhys}. 
The spatial distribution of Mn$^{4+}$ and that of the electric field are very similar to those of the metal-semiconductor junction except for the very small spatial scale.

\section{Conclusion}

The interfacial structures of two samples with different stacking orders, LNO/LMO/STO substrate (NMT) and LMO/LNO/STO substrate (MNT), were examined by precise atomic displacement and Mn valence distribution measurements. The interplane distance of the LMO region in MNT was very close to the reported lattice spacing of an as-grown LMO film\cite{Choi09JPhysD} with ferromagnetic ordering. In contrast, the interplane distance of LMO in the NMT sample was between those of annealed and as-grown LMO films. Using the lattice spacing, the valence of Mn ions was estimated as Mn$^{3.12+}$ and Mn$^{3.19+}$ for NMT and MNT, respectively. The obtained valence is quantitatively consistent with the values derived by the energy spectra of the CTR signal.

Local polarization was examined by the displacement of the A-site ions. Polarization was found in the LMO region for MNT and in the LMO/LNO interface region for NMT. The magnitude of the local field estimated by the La displacement was approximately 1~GV/m, which is a typical value for perovskite oxide heterointerfaces.
The local electric field pointed from LMO to LNO regardless of the stacking order. Our simultaneous observation of the valence distribution and detailed atomic arrangement demonstrates the correlation between the local electric field and the Mn$^{4+}$ distribution, which corresponds to the space charge layer.

\begin{acknowledgments}

The authors are grateful to Prof. H.~Shimotani for helpful discussions.
This work was supported by the MEXT Elements Strategy Initiative to Form Core Research Center (Grant No. JPMXP0112101001). The synchrotron radiation experiments were performed with the approval of the Photon Factory Program Advisory Committee (Proposal No.~2015S2-009). We thank Edanz Group (https://en-author-services.edanz.com/ac) for editing a draft of this manuscript.

\end{acknowledgments}

%

\end{document}